\documentclass[12pt]{iopart}

\usepackage{iopams}

\usepackage[numbers]{natbib}
\usepackage{graphicx,color}
\usepackage{hyperref}
\usepackage[dvipsnames]{xcolor}
\usepackage{enumitem}

\hypersetup{
  colorlinks=true,        % false: boxed links; true: colored links
  linkcolor=blue,         % color of internal links                     
  citecolor=cyan          % color of links to bibliography              
}

\usepackage{ragged2e}
\usepackage{aas_macros}                                                            

\graphicspath{{./plots/}}                         
\newcommand{\eg}{e.g.,~}
\newcommand{\eqref}[1]{(\ref{#1})}

\begin{document}

\title[Towards asteroseismology of NSs with PINNs]{Towards asteroseismology of neutron stars with physics-informed neural networks}

\author{Dimitra Tseneklidou}
\address{Departamento de Astronom\'{\i}a y Astrof\'{\i}sica, Universitat de Val\`encia,
  Dr. Moliner 50, 46100, Burjassot (Val\`encia), Spain}
  \ead{dimitra.tseneklidou@uv.es}

\author{Alejandro Torres-Forn\'e}
\address{Departamento de Astronom\'{\i}a y Astrof\'{\i}sica, Universitat de Val\`encia,
  Dr. Moliner 50, 46100, Burjassot (Val\`encia), Spain}                       
  \address{Observatori Astron\'omic, Universitat de Val\`encia,
C/ Catedr\'atico Jos\'e Beltr\'an 2, 46980, Paterna (Val\`encia), Spain}   
                     
\author{Pablo Cerd\'a-Dur\'an}
\address{Departamento de Astronom\'{\i}a y Astrof\'{\i}sica, Universitat de Val\`encia,
  Dr. Moliner 50, 46100, Burjassot (Val\`encia), Spain}
\address{Observatori Astron\'omic, Universitat de Val\`encia,
C/ Catedr\'atico Jos\'e Beltr\'an 2, 46980, Paterna (Val\`encia), Spain}

\vspace{10pt}
\begin{indented}
\item[]April 2025
\end{indented}

\begin{abstract}

The study of the gravitational wave signatures of neutron star oscillations may provide important information of their interior structure and Equation of State (EoS) at high densities. We present a novel technique based on physically informed neural networks (PINNs) to solve the eigenvalue problem associated with normal oscillation modes of neutron stars. The procedure is tested in a simplified scenario, with an analytical solution, that can be used to test the performance and the accuracy of the method. We show that it is possible to get accurate results of both the eigenfrequencies and the eigenfunctions with this scheme. 
The flexibility of the method and its capability of adapting to complex scenarios may serve in the future as a path to include more physics into these systems. 

\end{abstract}

%\maketitle

%%%%%%%%%%%%%%%%%%%%%%                                                                                
%%%  Introduction  %%%                                                                                
%%%%%%%%%%%%%%%%%%%%%%                                                                                 
\section{Introduction}
\label{sec:Introduction}
The asteroseismology of neutron stars (NSs) tries to understand the internal structure and composition of these objects by analyzing the imprint of global oscillations modes in their electromagnetic or gravitational wave (GW) emission. It is also a unique way of constraining the Equation of State (EoS) of matter at high densities, in a regime inaccessible in the laboratory. The field has been gaining increasing interest, 
after the detection of gravitational waves (GWs) from a binary neutron star merger along with concurrent electromagnetic observations \cite{GW170817_ligo}.
This single simultaneous observation set constraints on the EoS of the NS \cite{LigoVirgo2018:EoS, 2018PhRvL.120q2703A} based on the measured orbital parameters of the system. Attempts were made to detect oscillation modes in the post-merger evolution with no success \cite{PostMerger2017}. However, in a similar event in the future, oscillations may be detectable when the advanced detectors, Advanced LIGO \cite{LIGOScientific:2014pky}, Advanced Virgo \cite{VIRGO:2014yos} and KAGRA \cite{Akutsu2021},  reach design sensitivity or with next-generation detectors, the Einstein Telescope \cite{ET2025} and Cosmic Explorer\cite{CE2021}.
In this scenario, the detection of oscillation frequencies could constrain the bulk properties of the remnant NS using quasi-universal relations \cite{YagiYunes2013,TopolskiRezzolla2024ApJ}. This information is complementary to the one obtained from the orbital parameters of the binary before merger. 

A different scenario where GW asteroseismology has a strong interest in the case of core-collapse supernovae. In this case
the excitation of the normal modes of oscillation of the proto-neutron star and the standing accretion shock
dominate the GW signal. 
Asteroseismology can link these oscillation frequencies with properties of the star, such as mass and radius. Using data from numerical simulations, it is possible to find EoS-independent universal relations \cite{Sotani2017, Torres_et_al_letter, Sotani2019} that relate the observed frequencies to the bulk properties of the star. 

The oscillations of NSs have been studied extensively using linear perturbation theory in simplified scenarios \cite{McDermott1983, Reisenegger1992, Ferrari:2002ut, Passamonti2005, Dimmelmeier2006, Kruger:2014pva,Camelio:2017nka}. Previous works in the framework of core-collapse supernovae (CCSNe) can be found in \cite{Sotani:2016uwn, Torres_et_al_II, Morozova:2018glm, 2025Advection}. The latter study implements spectral methods, while the other ones apply common iterative integrators, like shooting method. These integrators have the disadvantage that they cannot be extended trivially to account for more complexity, \eg rotation and thus more realistic/physical scenarios. Our target is to solve the generalized eigenvalue problem related to the oscillation modes of a NS. We pursue a method that efficiently identifies all the eigenfrequencies. In this attempt, we focus on a machine learning approach, the physics-informed neural networks (PINNs).

In recent years, PINNs \cite{Raissi:2019} have been widely adopted for solving systems of partial differential equations (PDEs). The universal approximation theorem \cite{Cybenko:1989, Hornik:1991} establishes that neural networks (NNs) can approximate any continuous PDE, allowing these networks to effectively solve such equations numerically. This capability arises primarily because neural networks encode PDE information within a loss function, which is minimized throughout the training process. Typically, this loss function simultaneously incorporates the underlying physical equations, along with their initial and/or boundary conditions.

The PINN approach solves the PDE system by computing both the solution and its derivatives using automatic differentiation. Automatic differentiation, particularly through backpropagation, is commonly used to calculate gradients of machine learning models with respect to their parameters, facilitating effective training through optimization methods \cite{Baydin:2015}. Here, the same technique computes derivatives of the model outputs with respect to the input coordinates, allowing the definition of a loss function based on the residual of the differential equation. Minimizing this loss thus directly corresponds to solving the differential equation. 

The main advantage of PINNs is the ease with which they can be implemented in user-friendly machine learning frameworks without the need for specific grids or special conditions on the equations beyond requiring solutions to be finite and regular. This mesh-free semi-analytical nature of PINNs enables flexible handling of complex geometries and boundary conditions and allows seamless integration of physical constraints directly into neural network training \cite{Karniadakis:2021}.

Specifically, PINNs have been increasingly applied to eigenvalue problems, demonstrating several advantages compared to traditional numerical methods, such as finite-difference, finite-element, or spectral solvers. PINNs inherently produce analytically differentiable solutions, providing smooth eigenfunctions that can be evaluated at any point within the domain. They also offer flexibility in incorporating physical constraints directly into the training process, such as normalization and orthogonality conditions, essential to capture multiple eigenmodes accurately.

However, applying PINNs to eigenvalue problems also faces notable challenges and drawbacks \cite{Wang:2021,Hao:2022, Yang:2023,Holliday:2023}. A major limitation is the substantial computational cost associated with network training, especially when addressing multidimensional or multimodal problems. Moreover, PINNs often exhibit convergence difficulties for higher-order eigenmodes due to the inherent spectral bias of neural networks, which prefer low-frequency, smoother solutions. Ensuring the accurate convergence of eigenvalues and eigenfunctions typically requires carefully designed loss functions, specialized training strategies such as sequential or iterative extraction of modes, and meticulous tuning of hyperparameters. These additional complexities can make the approach more cumbersome compared to traditional solvers, which generally have well-established and efficient numerical algorithms.

In the field of astrophysics, PINNs have shown promising results in determining complex eigenfrequencies, particularly in computing black hole quasi-normal modes, highlighting their potential for astrophysical eigenvalue problems \cite{Cornell:2022,Luna:2023}. Additionally, PINNs have successfully tackled other diverse astrophysical challenges, such as efficient modeling of neutron star magnetospheres \cite{Urban:2023} and robust simulations of radiative transfer \cite{Mishra:2021}. These applications emphasize the ability of PINNs to handle complex boundary conditions, achieve computational efficiency, and produce accurate solutions consistent with fundamental physical laws \cite{Dai:2024,Athalathil:2024}.

In this paper, we explore the application of PINNs to solve the eigenvalue problem associated to the calculation of the normal modes of oscillation of neutron stars, with interest in asteroseismology. As a proof or principle, we consider in this work the case of the linear perturbations of a sphere of constant density. This scenario, while simple, is a basic step to understand the possibilities of PINNs as numerical eigenvalue solvers to be used as an alternative to traditional methods in this field. Our goal is to explore the needs in terms of network architecture and the loss requirements to find, not only the correct eigenfrequencies, but also the eigenfunctions with the precision required to conduct and study like the ones presented in \cite{Torres_et_al_I, Torres_et_al_II}. 

The paper is structured as follows: In Sec. \ref{sec:equations} we present the eigenvalue problem that we are going to use as a test.
In Sec. \ref{sec:PINN}, we explain how PINNs operate and describe in detail our implementation, both in concept and architecture. 
Sec. \ref{sec:results} is dedicated to the results. Here, we discuss thoroughly the convergence of the PINNs, the eigenvalues and the eigenfunctions obtained, as well as the choice of our hyperparameters. Lastly, we explain our main conclusions of this work in Sec. \ref{sec:conclusions}.

%%%%%%%%%%%%%%%%%%%%%%%%%%%%%%%%%     
%%%  The equations - maths  %%%                                   
%%%%%%%%%%%%%%%%%%%%%%%%%%%%%%%%%   
\section{Linear perturbations of a sphere of constant density}
\label{sec:equations}

As a simple test for the PINN eigenvalue solver, we consider the case of the normal oscillation modes of a sphere of unit radius, with constant density, pressure, and sound speed. For simplicity, we take these three quantities equal to unity. This test case was originally presented in the Appendix A of \cite{Torres_et_al_I}. Considering linear adiabatic perturbations of the Euler equations describing the dynamics of the fluid, it is possible to write the resulting system as an eigenvalue problem. In this work, we follow a procedure analogous to the formalism developed in \cite{2025Advection}.

Let us introduce the Lagrangian displacement, $\xi^i$, which expresses the displacement of a fluid element with respect to its position at rest. For a spherically symmetric background, it is possible to expand the displacement as
\begin{eqnarray}
    \xi^r &= \eta_1 Y_{lm} e^{-i\sigma t} , \label{eq:xi_r} \\
    \xi^{\theta} &= \eta_2\frac{1}{r}\partial_{\theta}Y_{lm} e^{-i\sigma t} ,\label{eq:xi_theta} \\
    \xi^{\phi} &= \eta_2\frac{1}{r\sin^2{\theta}}\partial_{\phi}Y_{lm} e^{-i\sigma t} , \label{eq:xi_phi}
\end{eqnarray}
where $\eta_1$ and $\eta_2$\footnote{Note that $\eta_1$ and $\eta_2$ relate to  $\eta_r$ and $\eta_{\perp}$ in \cite{Torres_et_al_I} as $\eta_1 = \eta_r$ and $\eta_2 = \eta_{\perp}/r$.} are two functions of $r$, related to the radial and angular amplitude of the Lagrangian displacement. We use spherical coordinates $\{r,\theta,\phi\}$, with $r\in[0,1]$. $Y_{lm}$ are the spherical harmonics and $\sigma$ is the oscillation frequency. We consider hereafter the case $l=2$, since it is the most relevant case for gravitational wave emission.

The perturbed Euler equations constitute a system of five partial differential equations that, for this particular case, can be reduced to the following system of two equations (see \citep{Torres_et_al_I,2025Advection}):
\begin{eqnarray}
    -r^2\partial_r^2\eta_1 -2r\partial_r \eta_1 +6r\partial_r\eta_2 +2\eta_1 -6\eta_2 &= r^2\sigma^2 \eta_1 \; , \label{eq:equation1} \\
    -r\partial_r\eta_1 -2\eta_1 +6\eta_2 &= r^2\sigma^2 \eta_2 \; .\label{eq:equation2}
\end{eqnarray}

At the origin we impose regularity of the displacement, which results in the boundary condition (see \cite{Torres_et_al_I})
\begin{equation}\label{eq:BC_origin}
  \eta_1(r=0)=\eta_2(r=0)=0.  
\end{equation}
At the outer boundary we impose
\begin{equation}\label{eq:BC_outer}
    \eta_1(r=1) = 0  \: .
\end{equation}

The aforementioned system of equations, together with the boundary conditions, represents a generalized eigenvalue problem, with $\sigma$ being the eigenvalues and $\eta_1$ and $\eta_2$ the eigenfunctions. The analytical solution of this problem reads
\begin{eqnarray}\label{eq:anal_solution_Bessel}
    \eta_1 &=& \eta_0 \partial_r\left[ j_l \left( r\sigma_n \right) \right] , \\
    \eta_2 &=& \frac{\eta_0}{r} j_l \left( r\sigma_n \right), 
\end{eqnarray}
where $j_l$ is the spherical Bessel function of the first kind and $\eta_0$ an arbitrary normalization constant fixing the amplitude of the mode. $\sigma_n$ is the eigenvalue of the $n$-th mode, that can be computed by imposing the boundary condition given by Eq.~\eqref{eq:BC_outer}, i.e. are the values of $\sigma_n$ such that
\begin{equation}
    \partial_r\left[ j_l \left( \sigma_n \right) \right]=0.
\end{equation}
The values for the first 6 eigenvalues can be found in table~\ref{tab:results}.

%%%%%%%%%%%%%%%%%%%%%%%%%%%%%%%%%%%%%%%%%%%%%%%%%%%%%%%%%  
\section{Physics Informed Neural Network}\label{sec:PINN}      
%%%%%%%%%%%%%%%%%%%%%%%%%%%%%%%%%%%%%%%%%%%%%%%%%%%%%%%%% 
PINNs \cite{Raissi:2019} employ neural networks to solve systems of PDEs by embedding the governing equations and boundary conditions directly into the training objective. They rely on the physical assumptions about the problem, which are incorporated into both the network architecture and the loss function. Modern machine learning frameworks \cite{Paszke:2019,Abadi:2016} enable automatic differentiation, allowing one to compute the derivatives of the network outputs. These represent the values of functions with respect to the input variables, which are the spatial coordinates. This allows to evaluate a system of PDEs in a domain, alongside with the corresponding boundary conditions. It is then possible to find solutions to the problem by optimizing the network parameters.

Consider a one-dimensional PDE defined over \(r \in [0,1]\) for an unknown function \(U(r)\). A generic form of such an equation is
\begin{equation}
{{\cal{F}}}\bigl(r;\,U(r),\,U'(r),\ldots;\,\Lambda\bigr) \;=\; 0,
\label{eq:PDE_1D}
\end{equation}
where \(\Lambda\) may represent additional parameters and the prime denotes derivatives with respect to $r$. 
We impose boundary conditions of the form
\begin{equation}
{\cal{B}}_{1}\!\bigl(U(0)\bigr)\;=\;0,
\quad
{\cal{B}}_{2}\!\bigl(U(1)\bigr)\;=\;0,
\label{eq:BC_1D}
\end{equation}
In the PINN approach, \(U(r)\) is replaced by a neural-network approximation \(\tilde{U}(r, \Theta)\), where $\Theta$ represents all the trainable parameters of the network (e.g. weights and biases). Substituting \(\tilde{U}\) in \eqref{eq:PDE_1D} defines the residual
\begin{equation}
{\cal{R}}(r;\,\Theta)
\;=\;
{\cal{F}}\bigl(r;\,\tilde{U}(r;\Theta),\,\tilde{U}'(r;\Theta),\ldots;\,\Lambda\bigr),
\label{eq:PDE_res_1D}
\end{equation}
which is a point-to-point measure of how well $\tilde U$ satisfies the PDE, for a particular set of parameters $\Theta$. A global measure, or PDE loss, can be constructed as 
\begin{equation}
    {\cal{L}} (\Theta)= \frac{1}{N_{\cal{R}}}\sum_{i=1}^{N_{\mathcal R}}||{\cal{R}}(r_i;\,\Theta)||^2,
\end{equation}
being $\{r_i\}$ a discrete set of $N_{\mathcal R}$ values of $r$ within the domain.
Minimizing \({\cal{L}}(\Theta)\) through gradient-based optimization drives \(\tilde{U}(r;\Theta)\) to satisfy \eqref{eq:PDE_1D}. %and \eqref{eq:BC_1D}.

Additionally, one has to impose the boundary conditions, Eq.~\eqref{eq:BC_1D}. 
There are two possible strategies. 
With soft enforcement, one adds an additional term to the loss depending on the boundary condition, using weights that balance the necessity to satisfy the PDE and the boundary conditions simultaneously. Additional terms may be included using the same procedure if further data (e.g., experimental measurements) are available.
Alternatively, one can use hard enforcement \cite{Lu:2021}, to ensure that $\tilde U$ satisfies the boundary conditions exactly by imposing analytical constraints or forcing the network architecture to respect them exactly. We use the latter approach in this work.
 
In the following sections, we adapt the PINNs methodology to address the case of the linear perturbations of a sphere of constant density, described above,
using residual definitions tailored to \(\eta_{1}(r)\) and \(\eta_{2}(r)\) along with their respective boundary conditions.

%%%%%%%%%%%%%%%%%%%%%%%%%%%%%%%%%     
%%%  The code  %%%                                          
%%%%%%%%%%%%%%%%%%%%%%%%%%%%%%%%%
\subsection{Eigenvalue Search}
\label{subsec:codeStructure} 
%%%%%%%%%%%%%%%%%%%%%%%%%%%%%%%%%
We aim at solving the eigenvalue problem, presented in \ref{sec:equations}, using PINNs. The solution of that problem is a set of infinite eigenvalues, but for practical applications, only the first few eigenmodes are needed. Therefore, it is not sufficient that the PINN finds a valid solution, but we need a procedure to systematically find all the eigenvalues within a frequency domain of interest, $\sigma \in [\sigma_{\rm min}, \sigma_{\rm max}]$. Here, the frequency, $\sigma$, must be considered as a parameter, which will be equal to the eigenvalue for particular values, for which the output of the network fulfills the conditions of the eigenvalue problem.
To find these particular values, we structure our eigenvalue search in two steps.
The first step consists of a {\it coarse search} along the domain of frequencies we want to analyze. For this purpose, we divide the frequency domain into a set of $n$ equally spaced intervals, $[\sigma_k, \sigma_{k+1}]$ with $k=0, ..., n-1$,
being $\sigma_0=\sigma_{\rm min}$ and $\sigma_n=\sigma_{\rm max}$, the boundaries of the frequency domain where the eigenvalues are searched.
In this coarse search, we determine if there is an eigenvalue within each interval.
In the second step, we perform the {\it{eigenfrequency computation}}, within each of the intervals that we determined contain a solution.

{\it{Coarse search:}} for each of the frequencies $\sigma_k$ we train a PINN, designed to solve the system of Eqs. \eqref{eq:equation1} and \eqref{eq:equation2}, together with the boundary conditions at $r=0$ (but not at $r=1$), using $\sigma$ as a variable parameter (non-trainable) and $\eta_1$ and $\eta_2$ as outputs. We refer to this network as s-PINN, hereafter. The eigenfrequencies of the system correspond to those values of the input $\sigma$, such that the output functions fulfill the boundary condition at $r=1$, i.e. $\eta_1(r=1)=0$.

\begin{figure}[t]
    \centering
    \includegraphics[width=0.8\linewidth]{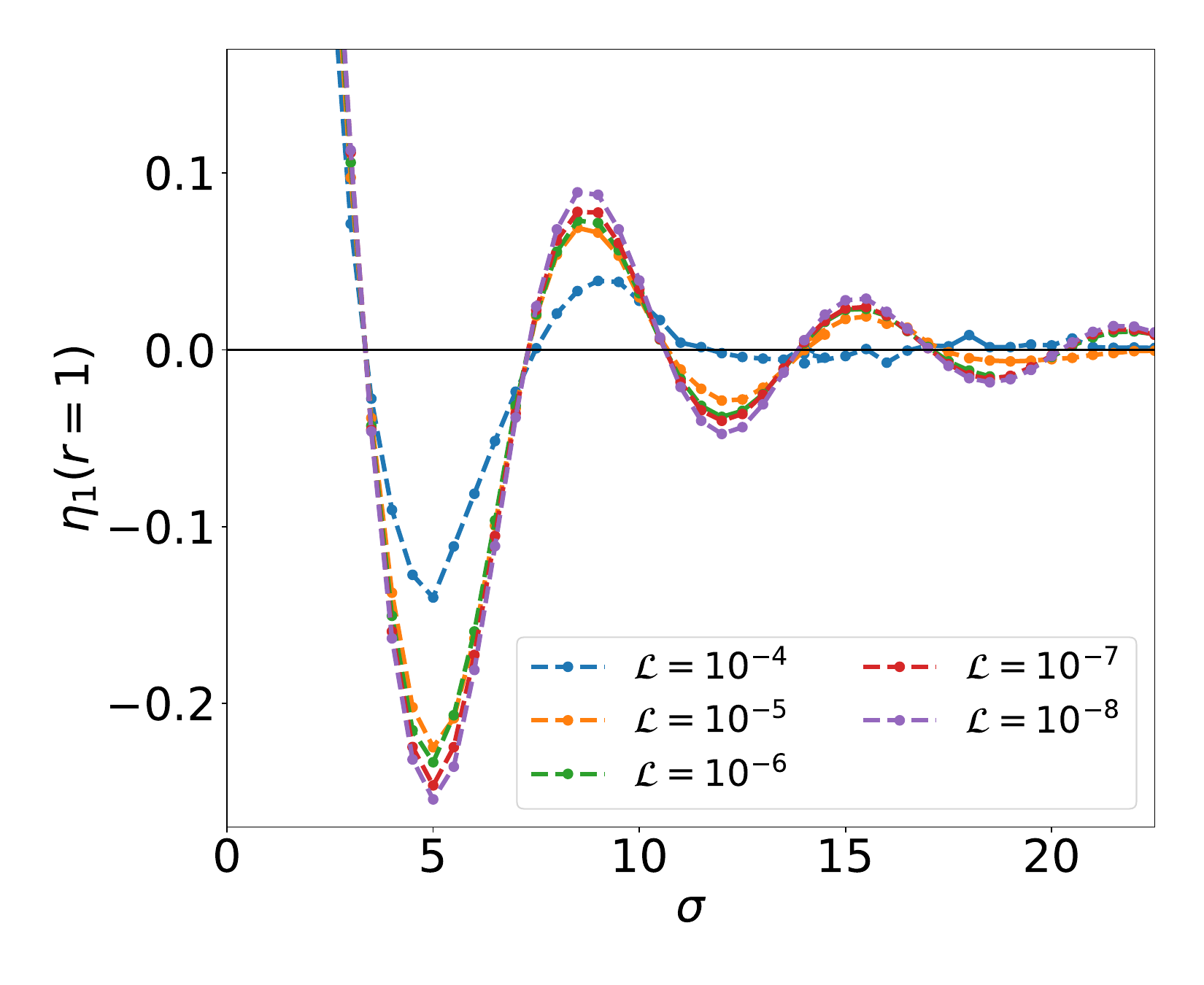}
    \caption{\justifying The figure shows the values of $\eta_1$ at the outer boundary $r=1$ with respect to the frequency. The frequencies that correspond to $\eta_1 \, \vline_{\:r=1}=0$ are the wanted eigenfrequencies.}
    \label{fig:intervals_roots}
\end{figure}

Fig.~\ref{fig:intervals_roots} shows the resulting value of $\eta_1(r=1)$, depending on the input frequency $\sigma$, evaluated on the coarse grid of frequencies, $\sigma_k$.
Wherever there is a change in the sign of $\eta_1 (r=1)$, then it is evident that within that interval, there exists a solution, and the interval is marked for detailed eigenfrequency calculation.

{\it Eigenfrequency computation:} once we have determined that a certain interval, $[\sigma_k, \sigma_{k+1}]$, contains an eigenvalue, we have developed two distinct methods to determine the eigenfrequency. In the first method, we use the same s-PINN, to search for the zero of $\eta_1 (r=1)$ (as a function of $\sigma$) within the interval, by means of a bisection algorithm. In the second method, we use a specific network for the computation, called f-PINN, hereafter. In this PINN, the boundary conditions at $r=1$ are incorporated, and $\sigma$ is a parameter of the network, instead of an input. The parameter $\sigma$ is initially set to $(\sigma_k + \sigma_{k+1})/2$, and the training of the PINN leads to its computation as a result.

%%%%%%%%%%%%%%%%%%%%%%%%%%%%%%%%%
\subsection{Neural Network Architecture}
\label{subsec:NNArchitecture}
\begin{figure}[t]
    \centering
    \includegraphics[width=1\linewidth]{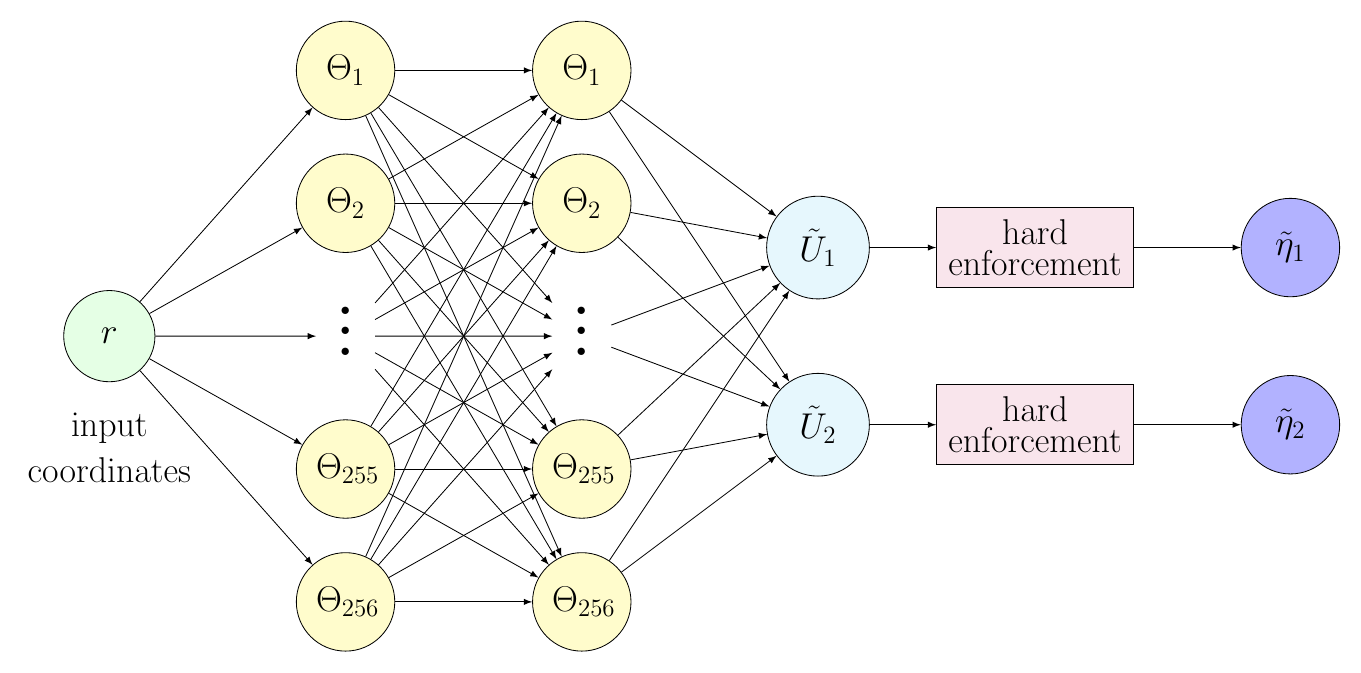}
    \caption{\justifying The cartoon depicts the s-PINN with one input node, $r$, and two output nodes, $\tilde{\eta}_1$ and $\tilde{\eta}_2$. The network consists of two hidden-layers with 256 neurons each.}
    \label{fig:pinn}
\end{figure}

%%%%%%%%%%%%%%%%%%%%%%%%%%%%%%%%%
Let us start the description of the architecture of the two PINNs used in this work, by presenting their common features. Both PINNs consist of a fully connected NN with 2 hidden-layers and 256 neurons each. The sole input node is the radial domain, $r$, while the two outputs are the eigenfunctions $\tilde{\eta}_1$ and $\tilde{\eta}_2$, as shown in Fig.~\ref{fig:pinn}. 

The weights are initialized through the Xavier function \cite{Glorot:2010}, and we use the Adam optimizer. The main difference between these two PINNs is that for the s-PINN, the frequency is a given value, while for the f-PINN, it is a parameter of the network. For the former, the activation function is the hyperbolic tangent, while for the latter, we use the sinusoidal. Although for both PINNs, the initial learning rate is $10^{-4}$, the f-PINN incorporates the ReduceLROnPlateau learning rate scheduler \cite{PyTorchReduceLROnPlateau}.
In Table \ref{tab:PINNs_characteristics} we summarize all the characteristics of each PINN, while in \ref{subsec:Hyperparameter} we explore the choice of the hyperparameters. Our code is written in Python and we use the PyTorch framework \cite{Paszke:2019} and its libraries.

\begin{table}[b]
    \centering
    \caption{\justifying Characteristics of the two PINNs in use: s-PINN and f-PINN.}
    \begin{indented}
    \item[]\begin{tabular}{@{}ccc}
    \br
         & s-PINN & f-PINN \\  \mr
      Hidden layers   & 2   & 2  \\ 
      Neurons   &  256    & 256  \\ 
      Optimizer    &  Adam    & Adam  \\
      Activation function  & Tangent &  Sinusoidal  \\
      Initialization of weights  &  Xavier  &  Xavier\\
      Learning rate  &  $10^{-4}$  &  $10^{-4}$\\
      LR scheduler &  -  & ReduceLROnPlateau\\
    \br
    \end{tabular}
    \end{indented}
    \label{tab:PINNs_characteristics}
\end{table}

%%%%%%%%%%%%%%%%%%%%%%%%%%%%%%%%%%%%%%%%%%%%%%%%%
\subsection{Loss Function}\label{subsec:loss_f}
%%%%%%%%%%%%%%%%%%%%%%%%%%%%%%%%%%%%%%%%%%%%%%%%%
The loss function of the PINN consists of the equations themselves. For that purpose, we recast the system of Eqs.~\eqref{eq:equation1} and \eqref{eq:equation2} to have the right-hand side of the equations equal to zero. Following the procedure outlined above, we substitute the solution of the equations, $\eta_1$ and $\eta_2$, by the numerical representation given by the output of the network, $\tilde{\eta}_1$ and $\tilde{\eta}_2$, resulting in equations for the residual functions,
$\mathcal{R}_1$ and $\mathcal{R}_2$, respectively, %according to
\begin{eqnarray}\label{eq:PDE=0}
    \left( C_{11}\partial_r^2 + C_{12}\partial_r + C_{13} + \sigma^2 C_{14}\right)\tilde{\eta}_{1}  
    + \left( C_{15}\partial_r   + C_{16}\partial_r \right)\tilde{\eta}_2  &= \mathcal{R}_1, \\
    \left( C_{21}\partial_r^2 + C_{22}\partial_r +C_{23}\right)\tilde{\eta}_1 
    + \left( C_{24}\partial_r  + C_{25} + \sigma^2 C_{26} \right) \tilde{\eta}_2 &= \mathcal{R}_2,
\end{eqnarray}
where %$\tilde{\eta}_1$ and $\tilde{\eta}_2$ are the outputs of the network, while 
$C_{ab}$, with $a=1,2$, representing each equation, and $b=1,...,N_C$, corresponds to the coefficients of the equations, being $N_C=6$, the number of coefficients in each equation. Note that this coefficients are in general functions of $r$.

It is straightforward to get the values of the coefficients by comparing the above equations directly with Eqs. \eqref{eq:equation1} and \eqref{eq:equation2}.
Starting with the s-PINN the steps we follow to calculate its loss are the following:

\begin{enumerate}

    \item We choose a {\it test grid} of $N_{\mathcal{R}}$ points $\{ r_i \}$
    within the domain of interest, $r \in [0,1]$. These points will be used as a reference to evaluate the residuals and thus play a role in the computation of the loss. Note however, that, since there is no numerical discretization in place, this grid does not play a role in terms of numerical resolution. Unless stated otherwise, we use $N_{\mathcal{R}}=100$ points, equally spaced in the domain.

    \item
    For each of the two equations, at each point of the test grid, we compute the norm
    \begin{equation}\label{eq:norm}
        \mathcal{N}_a (r_i) = \frac{\sqrt{\sum_{b=1}^{N_C} C_{ab}(r_i)^2}}{N_C} \quad ; \quad a=1,2,
    \end{equation}
    
    We use these quantities in the next step to normalize the residuals. 

    \item We compute the loss of each equation 
    as the mean squared value of the normalized residuals over the test grid
    \begin{equation}\label{eq:loss_PDE}
        \mathcal{L}_a = \frac{1}{N_{\mathcal{R}}} \sum_{i=1}^{N_\mathcal{R}} 
        \bigg|\bigg| \frac{\mathcal{R}_a (r_i)}{\mathcal{N}_a (r_i)} \bigg|\bigg|^2 \quad; \quad a=1,2,
    \end{equation}
    
    \item The total loss of the s-PINN is the the sum of the losses of each equation
    \begin{equation}\label{eq:loss_s-pinn}
        \mathcal{L}_s = \mathcal{L}_1 + \mathcal{L}_2.
    \end{equation}
    
\end{enumerate}

In the case of the f-PINN we follow the same steps as the s-PINN, but we add an additional step, as we want to force the PINN to find a solution within a confined interval $[\sigma_k, \sigma_{k+1}]$:
\begin{enumerate}[resume]
    \item The total loss of the f-PINN is then   
%    Thus in that case the loss will be
    \begin{equation}\label{eq:loss_f-pinn}
        \mathcal{L}_f = \mathcal{L}_s \cdot \frac{2}{\tanh\left(\frac{\sigma - \sigma_k}{0.01}\right)+1} \cdot \frac{2}{\tanh\left(\frac{\sigma_{k+1} - \sigma}{0.01}\right)+1}. \;
    \end{equation}
    \end{enumerate}
    Recall that for the f-PINN $\sigma$ is a parameter of the network.
    
The extra terms ensure that the loss becomes infinite at the boundaries of the interval, preventing the search to wander outside the interval. Inside the interval the extra factors are essentially $1$, so the loss remains unaffected within it.

The PINN solves the system of equations successfully, when it reaches the desired threshold value for the loss function, $\mathcal{L} \leq \varepsilon$. The choice of this value is discussed thoroughly in \ref{subsec:Hyperparameter}. In addition, achieving the minimum value of the loss is not the only criterion to admit a solution. We discuss this topic in detail in \ref{sec:results}.

%%%%%%%%%%%%%%%%%%%%%%%%%%%%%%%%%%%%%%%%%%%%%%%%%%%%%%%%%%%%
\subsection{Boundary Conditions}\label{subsec:boundaryConditions}
%%%%%%%%%%%%%%%%%%%%%%%%%%%%%%%%%%%%%%%%%%%%%%%%%%%%%%%%%%%%
Here, we present the numerical implementation of the boundary conditions, both at the origin and at the outer boundary, presented in \ref{sec:equations}. That is done in the form of hard enforcement, meaning that the solution is not represented by the output of the neural network, but by a function of the output, that ensures 
that the boundary conditions are satisfied. 
In our case, we use the following format for this function
\begin{equation}
  \tilde \eta_a(r) = \tilde{U}_a(r)g_a(r)+ {h_a}\left(r, \eta_a (r_{\rm boundary}), \eta^{\prime}_a(r_{\rm boundary}) \right),
\end{equation}
with $a=1,2$, and where $\tilde{U}(r)$ is the raw output of the network, and $g_a$ and $h_a$ are two auxiliary functions. Note that $h_a$ depends not only on $r$ but also on the value of the function and its derivatives at the boundary, $r_{\rm boundary}$, in general. Depending on the case $r_{\rm boundary}$ may refer to the inner or the outer boundary.

For the s-PINN we only impose boundary conditions at the inner boundary ($r=0$), and the output of the network, after the hard enforcement, reads
\begin{eqnarray}
   \tilde\eta_{1}(r) &=& r^2\, \tilde{U}_1(r)+ r\, \eta_1^\prime(0) + \eta_1(0),
   \label{eq:hard_enforc_Spinn_1} \\
    \tilde\eta_{2}(r) &=& r\, \tilde{U}_2(r)  + \eta_2(0) .
    \label{eq:hard_enforc_Spinn_2}
\end{eqnarray}
To ensure that Eq.~\eqref{eq:BC_origin} is fulfilled we use $\eta_1(0)=0$ and  $\eta_2(0)=0$. By choosing a value for $\eta_1^\prime(0)$ we fix the amplitude of the eigenfunction, which is arbitrary, and discard the trivial solutions of the system. Therefore, we impose $\eta_1^\prime(0) = 1$, although any other value would be valid in principle.
Note that, by construction, $\tilde \eta_a(r)$ fulfills exactly the same boundary conditions as $\eta_a(r)$ at $r=0$.

For the f-PINN, we impose both boundary conditions at $r=0$ and $r=1$, given by Eqs.~\eqref{eq:BC_origin} and \eqref{eq:BC_outer}, additionally to the normalization condition imposed by setting the value of $\eta'_1(0)$. After the hard enforcement, the output of the f-PINN reads
\begin{eqnarray}
    \tilde\eta_{1}(r) &=& \left[ \tilde{U}_1(r) \left( 1 - e^{-r} \right)^2 + \left( 1 - e^{-r} \right) \right] \frac{1 - e^{1-r}}{1 - e},
    \label{eq:eta_r_r_1}
\\
     \tilde\eta_{2}(r) &=& \tilde{U}_2(r) \left( 1 - e^{-r} \right).
     \label{eta_perp_r_1}
\end{eqnarray}
These functions automatically satisfy the boundary conditions, $\tilde\eta_1 (0)=\tilde\eta_2 (0)=\tilde\eta_1 (1)=0$,
and the normalization condition $\tilde\eta'_1(0)=1$.

For the f-PINN the hard enforcement is an exponential function in contrast to the s-PINN. The reason is that we found out that the exponential boundary conditions in combination with the sinusoidal activation function improves the rate of convergence of the loss. 
Different functions, such as a combination of an exponential and a polynomial function, have also been explored. However, the f-PINN was not converging to any solution.

%%%%%%%%%%%%%%%%%%%%%%%%%%%%%%%%%%%%%
\section{Results}\label{sec:results}
%%%%%%%%%%%%%%%%%%%%%%%%%%%%%%%%%%%%%
In this section, we show the results that we obtained by solving the eigenvalue problem. 
We present the numerically computed eigenvalues and eigenfrequencies for both methods, bisection and f-PINN, and compare them with the analytical ones, as well as to each other. Although the implementation of the code has been explained in detail in \ref{sec:PINN}, here we explore the different options for the hyperparameters and justify our choices. In addition, we include a study about the convergence of both PINNs.

%%%%%%%%%%%%%%%%%%%%%%%%%%%%%%%%%%%%%
\subsection{Convergence of the PINNs}
\label{subsec:convergence}
%%%%%%%%%%%%%%%%%%%%%%%%%%%%%%%%%%%%%
The PINN converges to a solution when it solves the system of equations \eqref{eq:equation1} and \eqref{eq:equation2}, with the corresponding boundary conditions, Eqs. \eqref{eq:BC_origin} and \eqref{eq:BC_outer}, up to a specified precision. That translates into the threshold value of the loss function, $\varepsilon$, to which the PINN has to reach. Once the loss function becomes equal or smaller to that value, $\mathcal{L}\leq \varepsilon$, then a solution has been found. Here, we discuss the importance of this threshold value and justify which value to use depending on the PINN.

In Fig.~\ref{fig:intervals_roots}, we show how the s-PINN, used during the first step of the search, is affected by the value of the loss function reached during training. We plot the computed values of $\eta_1$ at the outer boundary, $r=1$, depending on the value of $\sigma_k$ in the coarse frequency grid. Recall that the zeros of this function correspond to the eigenvalues, and that the coarse grid search serves to bound the eigenvalues within the intervals between two values of $\sigma_k$.
It is evident that the higher the loss, the less accurate the calculation of $\eta_1(r=1)$. Not only do the values of $\eta_1(r=1)$ have a larger deviation, but the intervals are misidentified. Thus, the inferred frequency during the second step of the search, whether using bisection or the f-PINN, will be incorrect. To avoid this problem we use a threshold of $10^{-6}$ in the coarse search, which we found it sufficient to determine the intervals containing the solution. When used in searching the eigenvalues, by means of the bisection method, we use instead a threshold of $10^{-7}$in the s-PINN. This allows to recover accurately the radial profiles of the eigenfunctions (see discussion below for the f-PINN).

\begin{figure}[t]
\centering
   \includegraphics[width=0.49\linewidth]{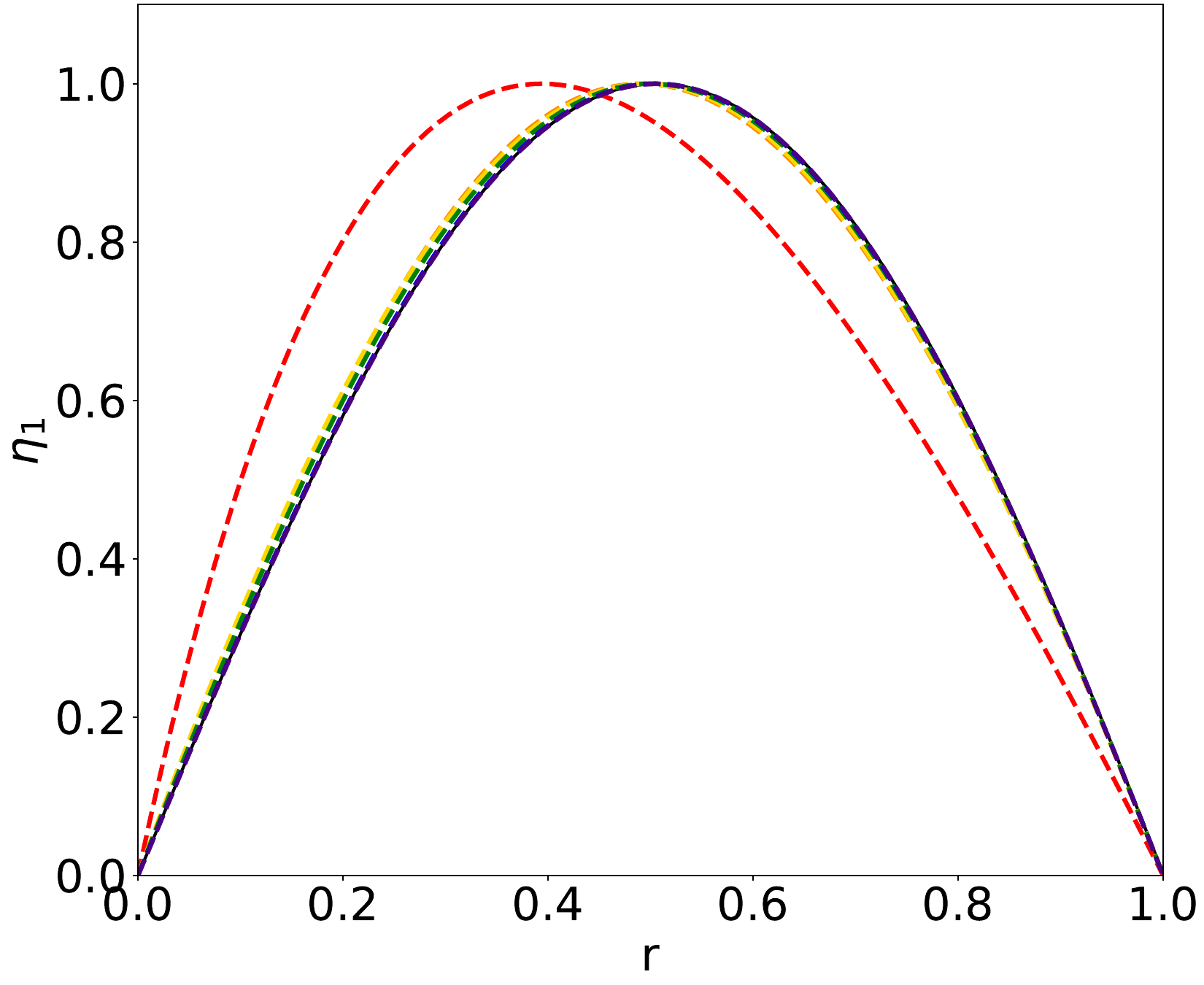} %\vspace{-1em}
   \includegraphics[width=0.49\linewidth]{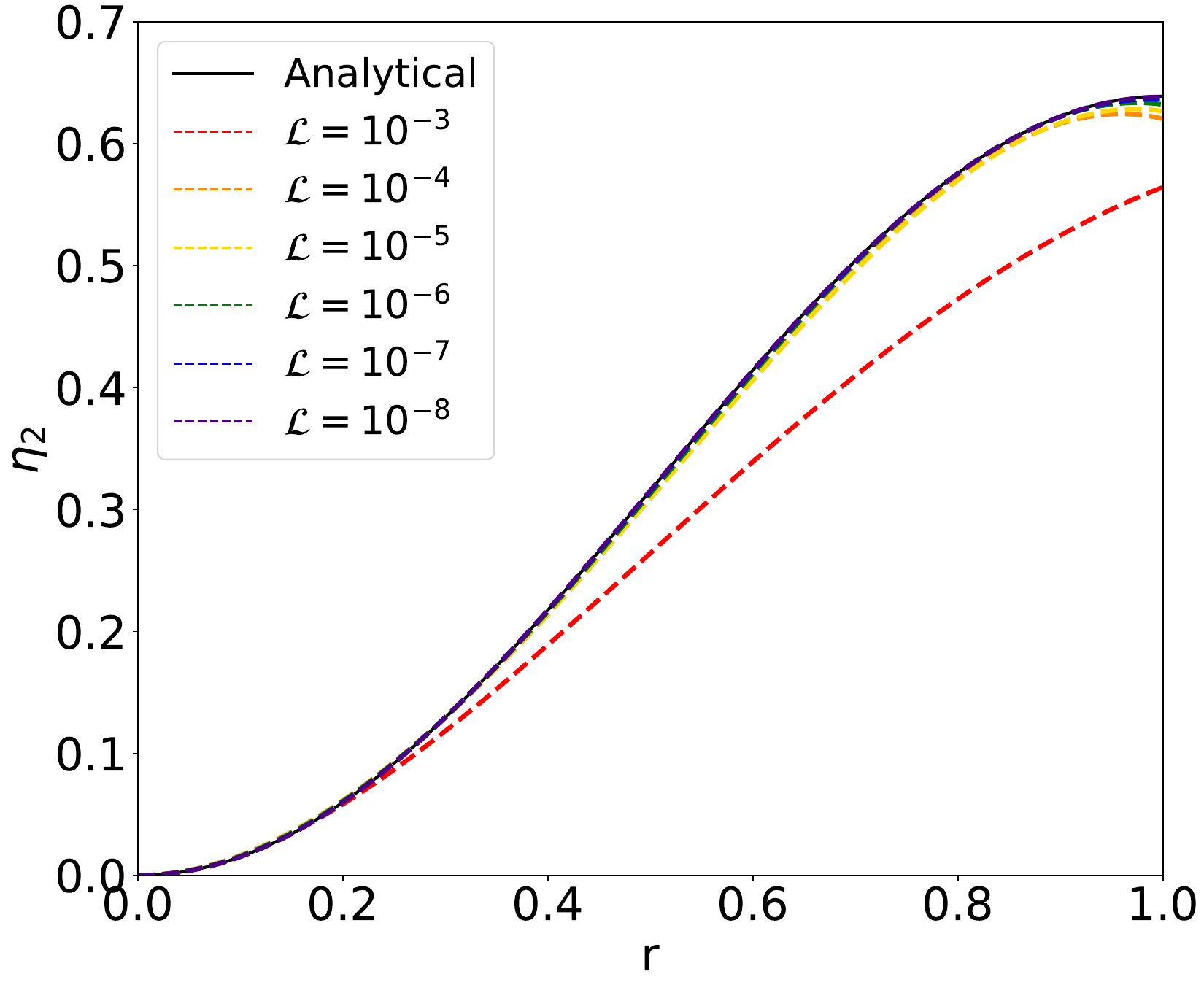}
\caption{\justifying The radial profiles of $\eta_1$ (upper figure) and $\eta_2$ (lower figure) for the first frequency, $\sigma_0 = 3.34$, for 256 neurons and different thresholds for the loss, using the f-PINN.}
\label{fig:eta_first_f_profiles_losses}
\end{figure}

\begin{figure}[t]
\centering
   \includegraphics[width=0.49\linewidth]{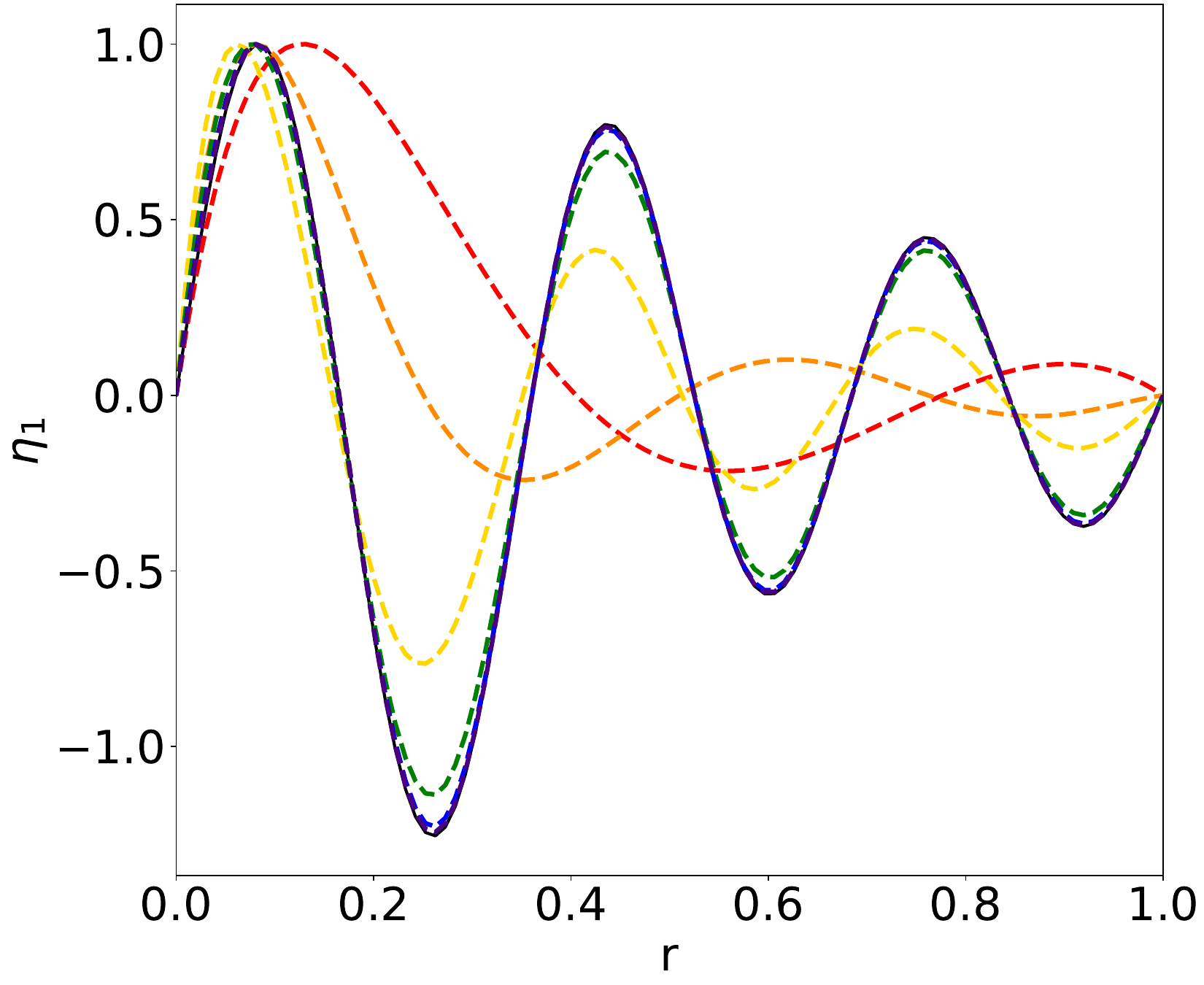}
   \includegraphics[width=0.49\linewidth]{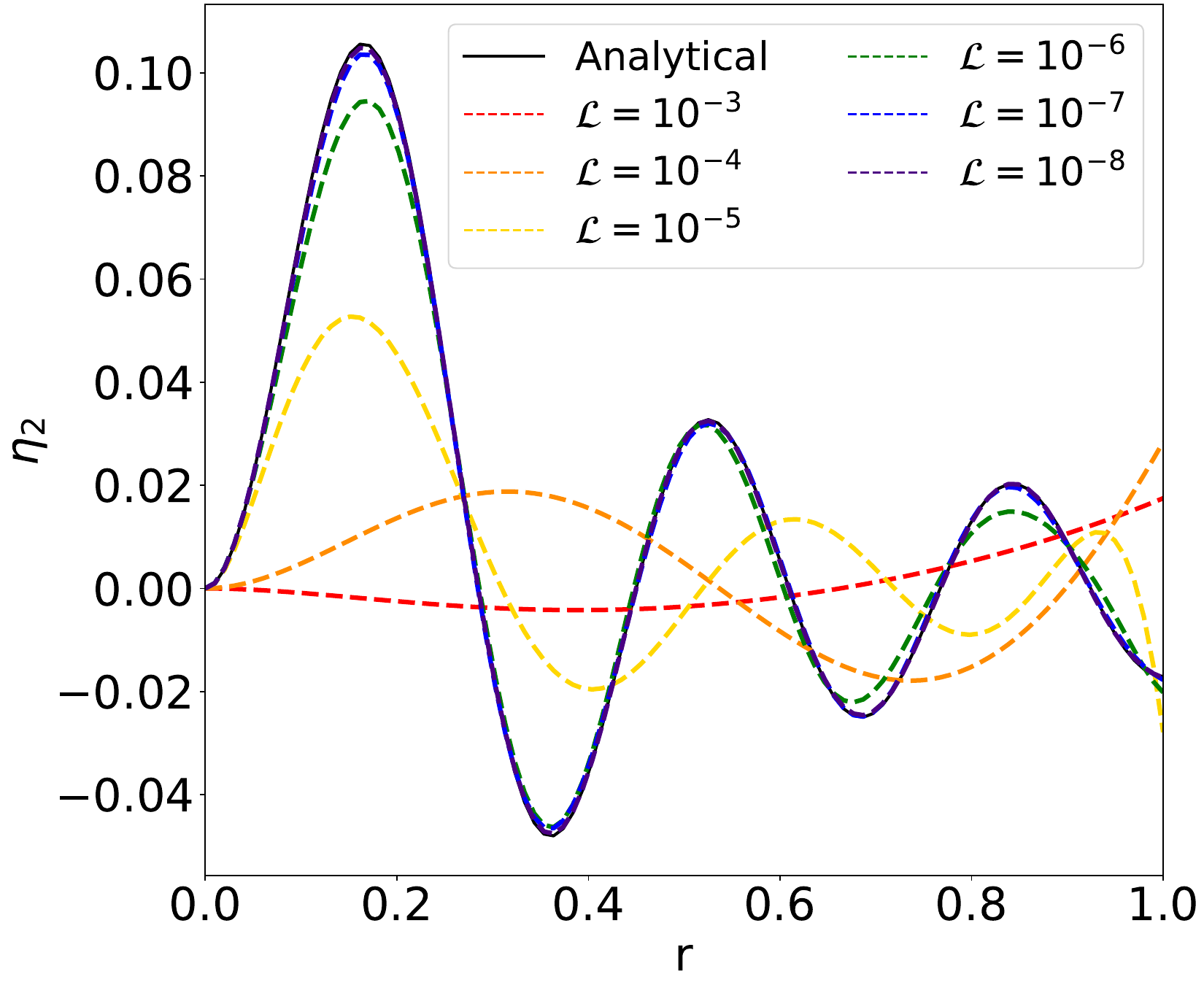}
\caption{\justifying The radial profiles of $\eta_1$ (upper plot) and $\eta_2$ (lower plot) for the higher-order harmonic, $\sigma_5=20.22$, for 256 neurons and different thresholds for the loss, using the f-PINN.}
\label{fig:eta_sixth_f_profiles_losses}
\end{figure}

To check the impact of the threshold value of the loss function for the f-PINN we plot in Figs. 
\ref{fig:eta_first_f_profiles_losses} and \ref{fig:eta_sixth_f_profiles_losses}
the eigenfunction of the lowest frequency mode, as well as of a higher-order mode (with five nodes), respectively, depending on the value of the loss. The number of neurons is kept constant and equal to 256. 
For the first mode (Fig.~\ref{fig:eta_first_f_profiles_losses}) we see that, even with a relatively large value for the threshold of $10^{-4}$, the solution is very close to the exact analytical solution. However, for the higher order harmonic (Fig.~\ref{fig:eta_sixth_f_profiles_losses}), a smaller threshold is needed to recover the analytical solution. 
Therefore, we chose a value of $10^{-7}$ for the threshold of the f-PINN, which should give good accuracy across a number of modes, at least the first few harmonics.

\begin{figure}[t]
\centering
   \includegraphics[width=0.49\linewidth]{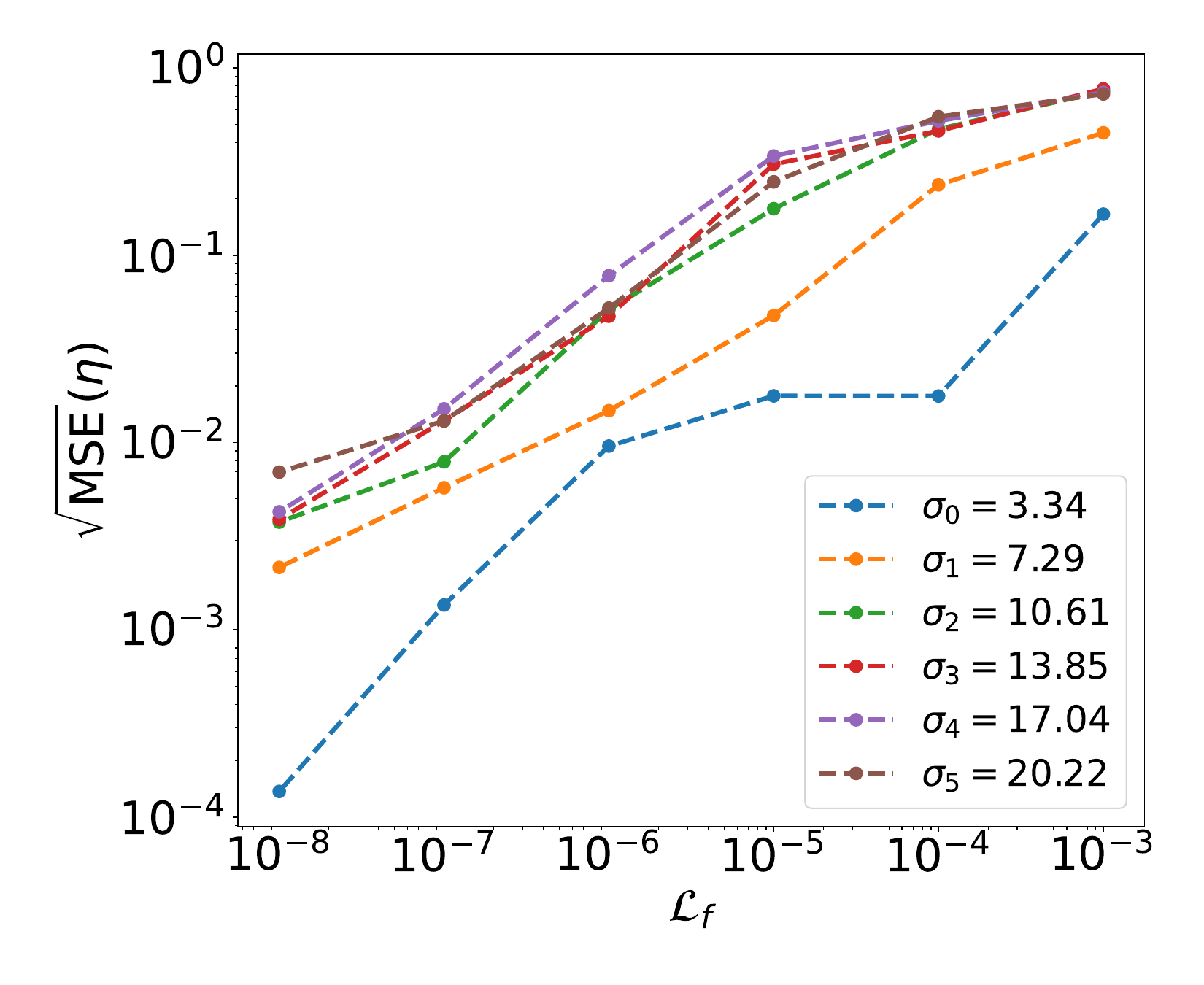}
   \includegraphics[width=0.49\linewidth]{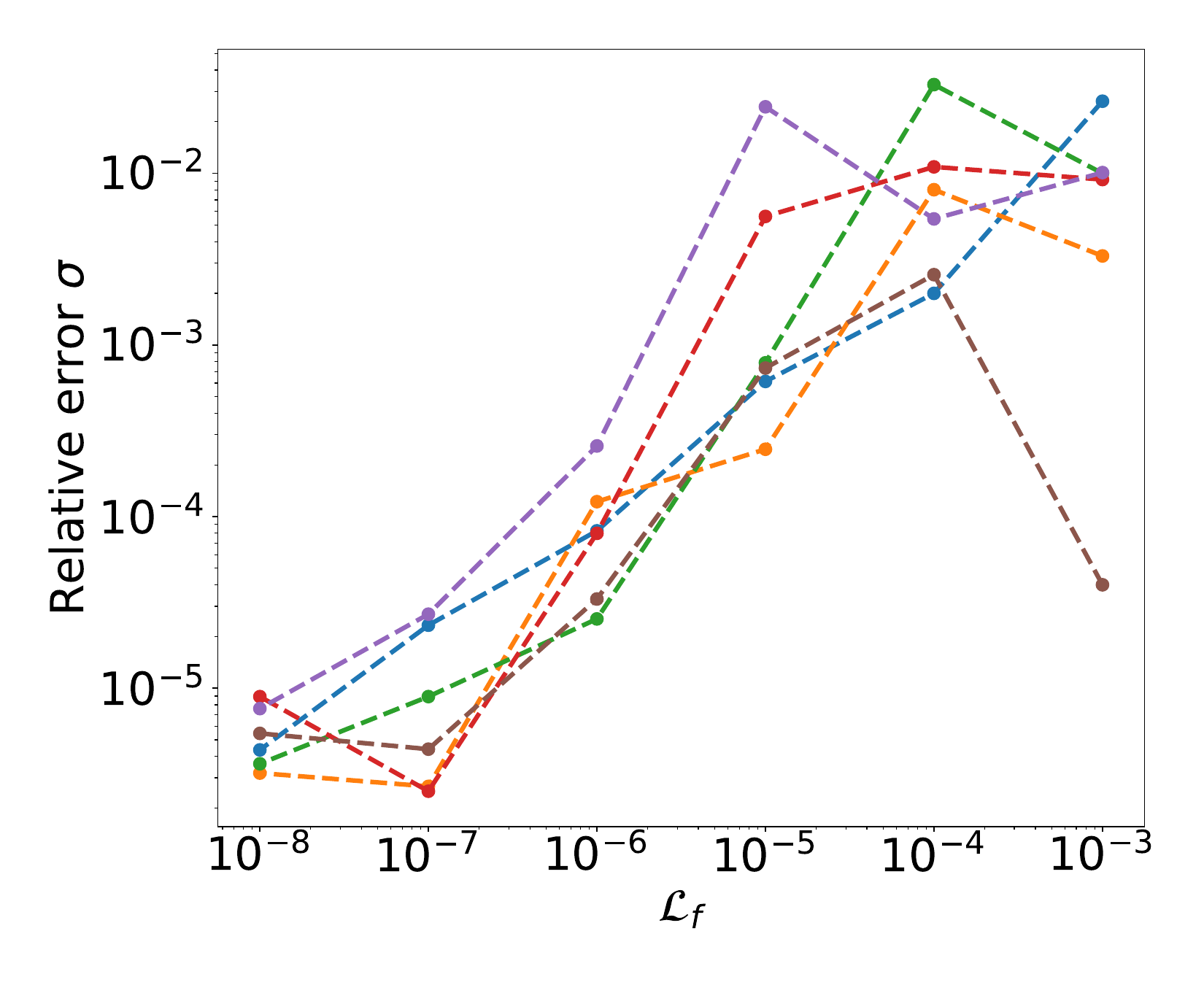}
  \caption{\justifying Upper panel: Mean square error of the variables $\eta$ across all eigenfrequencies for various values of the loss function, $\mathcal{L}_f$, using the f-PINN. Bottom panel: Relative error of the eigenfrequencies for these same loss values. The network has 256 neurons.}
   \label{fig:errors_all_losses} 
\end{figure}

In the upper figure of Fig.~\ref{fig:errors_all_losses}, we plot the square root of the sum of the mean square errors of the eigenfunctions, $\sqrt{{\rm MSE}(\eta_1) + {\rm MSE}(\eta_2)}$, with respect to the loss for the f-PINN, $\mathcal{L}_f$, with 256 neurons. As expected, the lower the loss, the lower the error. Recall that the loss function represents the system of the equations and thus the value $\varepsilon$ defines the accuracy to which they will be solved. The observed trend is dominated by the mean square error (MSE) of $\eta_1$, as it is always larger than for $\eta_2$ (see Table \ref{tab:results}). In the lower plot of Fig.~\ref{fig:errors_all_losses}, we show the relative error of the eigenfrequencies with respect to the loss function $\mathcal{L}_f$. Notice that although we can achieve a satisfying relative error of the order of $10^{-4}$ for $\varepsilon=10^{-6}$, the $\sqrt{\rm MSE}$ is two orders of magnitude larger than the relative error of the frequencies. The consequence is the mismatch of the eigenfunctions observed in Figs. \ref{fig:eta_first_f_profiles_losses} and \ref{fig:eta_sixth_f_profiles_losses}. Finally, note that for the frequency $\sigma_5=20.22$, the relative error can occasionally be low even for high loss values, due to the stochastic nature of the calculation.

Closing this section, it is worth pointing out that, unlike classical numerical methods (e.g., finite differences), where resolution explicitly depends on grid spacing, PINNs do not inherently depend on a structured grid. However, inaccuracies can still occur if the distribution or number of points in the test grid is insufficient, potentially leading to approximation errors. To rule out this possibility, we examined the results obtained with different numbers of points in the test grid. Specifically, for the s-PINN, we compared the results of two test grids with 100 and 200 equally spaced points, and observed no significant differences in the eigenfrequencies nor the eigenfunctions.
Consequently, achieving a solution with PINNs requires achieving a specific accuracy, which is directly related to the value of the loss function.

%%%%%%%%%%%%%%%%%%%%%%%%%%%%%%%%%%%%%%%%%%%%%%%%
\subsection{Solutions of the Eigenvalue Problem}
\label{subsec:Solutions}
%%%%%%%%%%%%%%%%%%%%%%%%%%%%%%%%%%%%%%%%%%%%%%%%
 We solve the eigenvalue problem applying a two-step method. First, the s-PINN searches the intervals containing a solution. For this case the search is performed in the region $\sigma \in [0.5, 23.0]$, with values of the coarse grid, $\sigma_k$ increasing in steps of $0.5$. Afterwards, we compute the eigenvalue and the eigenfunctions with two distinct methods; bisection and f-PINN. The bisection method uses the s-PINN as the PDEs solver for a given frequency. In contrast, the f-PINN has the frequency as a parameter of the network, while the initial guess is always the center of the interval. In this section, we present the results obtained from each of these methods.
As our system of equations has analytical solutions, we can calculate the relative error of the frequencies and the mean square error between the eigenfunctions obtained by the network and the theoretical ones.

In Table \ref{tab:results}, we show the first six eigenfrequencies calculated using the bisection method and the f-PINN. For comparison, we also provide the analytical values of the frequencies. For each method, we present the time needed to compute each eigenvalue and its relative error compared to the theoretical value. The relative error of the frequencies computed by the s-PINN is of the order of $10^{-4}$, while for the f-PINN, it is around two orders of magnitude lower.

Although the f-PINN achieves higher accuracy in deriving the eigenfrequencies, it is significantly slower compared to the bisection method. For comparison, the calculation of the first six eigenvalues, including the coarse grid search,  with the bisection method requires $1.40 \cdot 10^4$ s, while with the f-PINN it needs $2.50 \cdot 10^4$ s. 
This difference in computational time is probably related to the use of transfer learning in the case of the bisection. In this case, the same s-PINN that is used for the first step (finding the intervals) it is used for the bisection, albeit with a lower loss requirement. In such a manner, we reuse in the second step (via transfer learning) the network parameters previously trained for frequencies near the target eigenfrequency. 
However, for the f-PINN, we cannot transfer the parameters of the s-PINN, because the network architecture is different. In this case, the parameters of the f-PINN are initialized for each interval, so it needs longer time to find the solutions. 
In addition, we tried transfer learning by initializing the f-PINN with the pretrained weights once the first eigenfrequency was found, but this did not improve our results. Since the eigenfrequency is a network parameter, any change in its value alters the network configuration.

Lastly, the corresponding eigenfunctions are also compared to the analytical ones. We include in the table the square root of their mean square error. We find that the $\sqrt{\rm MSE}$ of the eigenfunctions is slightly improved for the f-PINN. The relative difference in the values between the two methods is of the order of $10^{-1}$.

\begin{table}[t]
    \centering
\caption{ Comparison between the results of the two methods searching for the eigenvalues: bisection and the f-PINN. For both of them, the corresponding PINN we used has 256 neurons and the threshold value of the loss function is $\varepsilon=10^{-7}$. For each case we show the numerically computed frequency, the computational time, the relative error of the frequency with respect to the analytic solution, and the square of the mean square error of the difference between the eigenfunction and its analytical value (separately for $\eta_1$ and $\eta_2$).}
    \begin{indented}
    \item[]\begin{tabular}{@{}ccccccc}\br
    &Analytical   &  Numerical & Time   & Relative error  & $\sqrt{\rm MSE}(\eta_1)$ & $\sqrt{\rm MSE}(\eta_2)$ \\ 
     & frequency & frequency & (s) & $(10^{-4})$ & $(10^{-3})$  & $(10^{-3})$  
      \\ \mr
     Bisection& 3.3421  & 3.3437 & 12.77 & 4.9 & 4.2 & 0.5 \\
      &7.2899  & 7.2891 & 4.35 & 1.2 & 7.6 & 3.5   \\
      &10.6138  & 10.6250 & 3.84 & 10.5 & 12.4 & 0.7   \\
      &13.8461  & 13.8750 & 12.66 & 20.9 & 15.0 & 1.2  \\
      &17.0429  & 17.0625 & 14.42 & 11.5 & 16.3 & 0.7   \\
      &20.2219  & 20.2500 & 6.56 & 13.9 & 18.3 & 1.3   \\
\mr
      f-PINN&3.3421   & 3.3422 & 4.30$\cdot 10^2$ & 0.2 & 1.2 & 0.6  \\
      &7.2899   & 7.2899 & 6.41$\cdot 10^2$ &  0.03 & 5.6 & 1.0  \\
      &10.6138  &  10.6139 & 2.17$\cdot 10^3$ & 0.09 & 7.8 & 0.7   \\
      &13.8461  &  13.8461 & 3.65$\cdot 10^3$ &  0.02 & 13.0 & 1.4 \\
      &17.0429  &  17.0424 & 8.31$\cdot 10^3$ &  0.3 & 15.0 & 1.4  \\
      &20.2219  &  20.2218 & 6.86$\cdot 10^3$ &  0.04 & 13.0 & 0.9  \\
    \br
    \end{tabular}
    \end{indented}
    \label{tab:results}
\end{table}

%%%%%%%%%%%%%%%%%%%%%%%%%%%%%%%%%%%%%
\subsection{Hyperparameter study}
\label{subsec:Hyperparameter}
\begin{figure}[t]
\centering
   \includegraphics[width=0.49\linewidth]{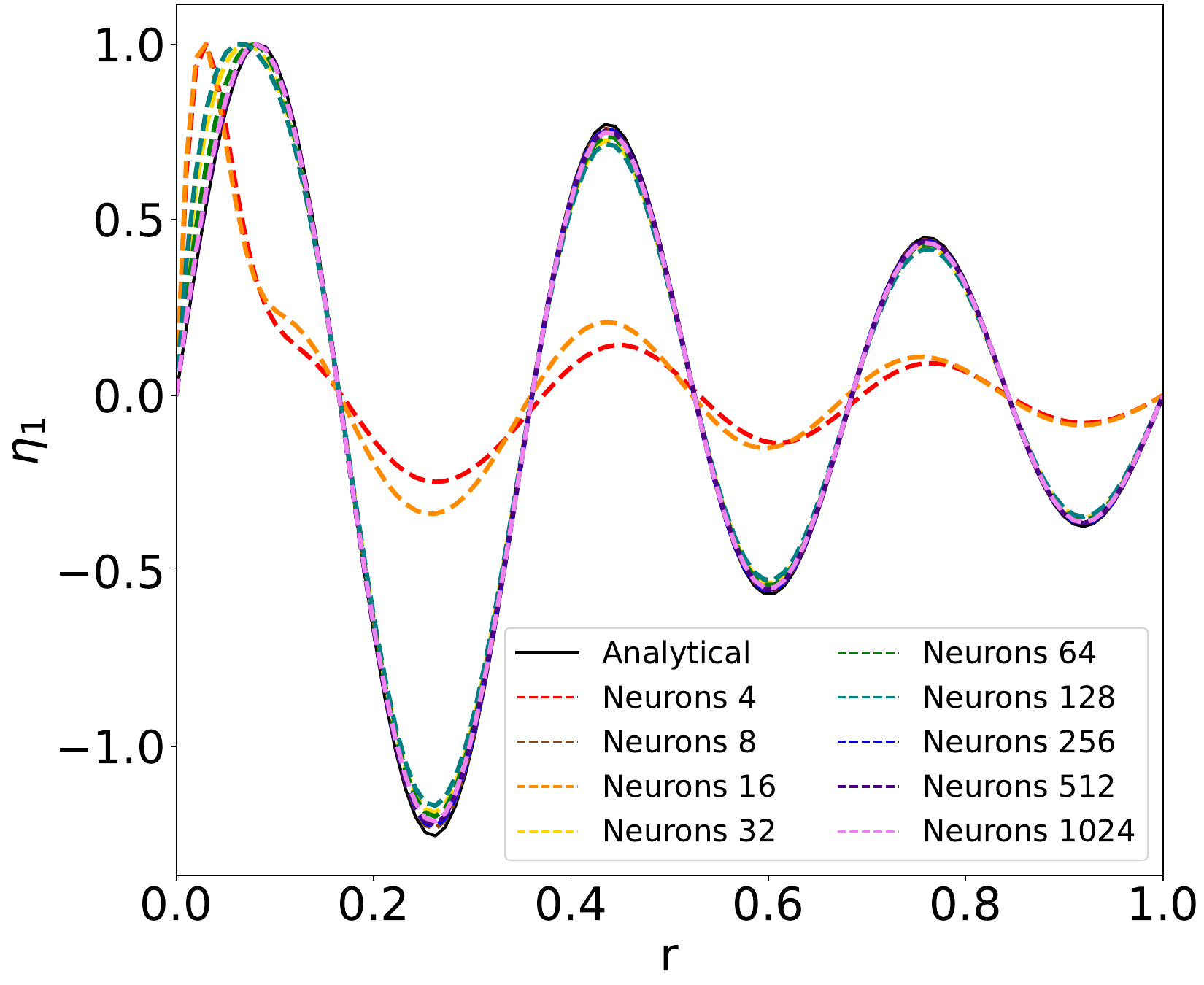}
   \includegraphics[width=0.49\linewidth]{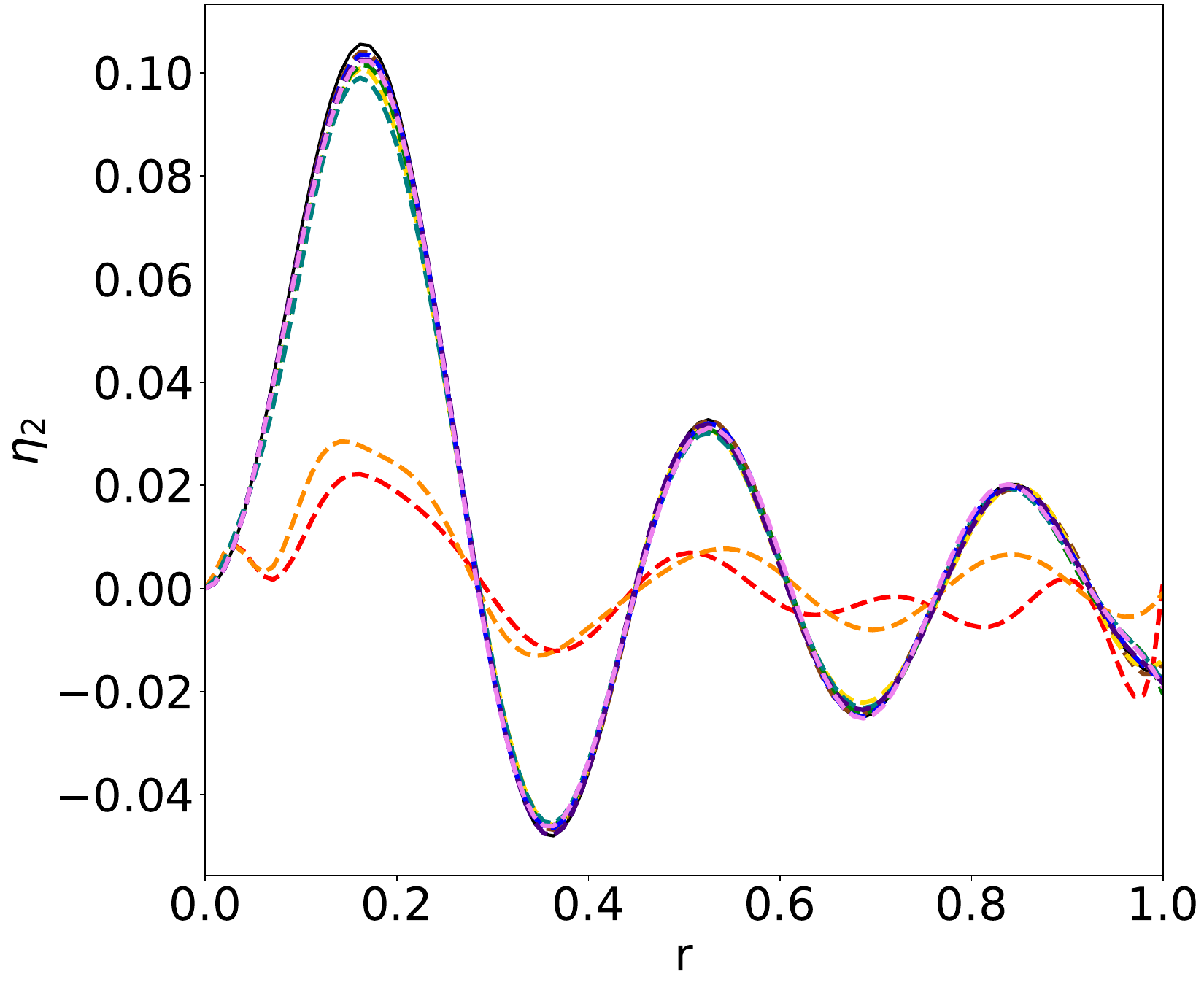}
\caption{Radial profiles of $\eta_1$ (upper panel) and $\eta_2$ (lower panel) for the higher-order mode with eigenvalue $\sigma_5=20.22$, for loss $10^{-7}$ for different number of neurons, computed with the f-PINN.}
\label{fig:eta_profiles_number_neurons}
\end{figure}

%%%%%%%%%%%%%%%%%%%%%%%%%%%%%%%%%%%%%
Our hyperparameter study begins by exploring network architecture features, specifically the number of hidden layers and neurons. We investigated networks with varying numbers of hidden layers, from 2 to 10, and observed that adding more layers did not significantly improve the accuracy of the solution. In addition, increasing the number of layers had a negative impact on computational time. Thus, we selected an architecture with 2 hidden layers.
The number of neurons plays a more important role, but mainly for the f-PINN as it affects the shape of the eigenfunctions, as shown in Fig.~\ref{fig:eta_profiles_number_neurons}. It is also worth mentioning that the number of neurons affects mostly the shape of the higher-order overtones, while the lowest frequency mode is computed precisely even for a low number of neurons. We choose 256 neurons for each layer to capture the complexity of the radial profiles of the eigenfunctions.

\begin{figure}[t]
   \includegraphics[width=0.49\linewidth]{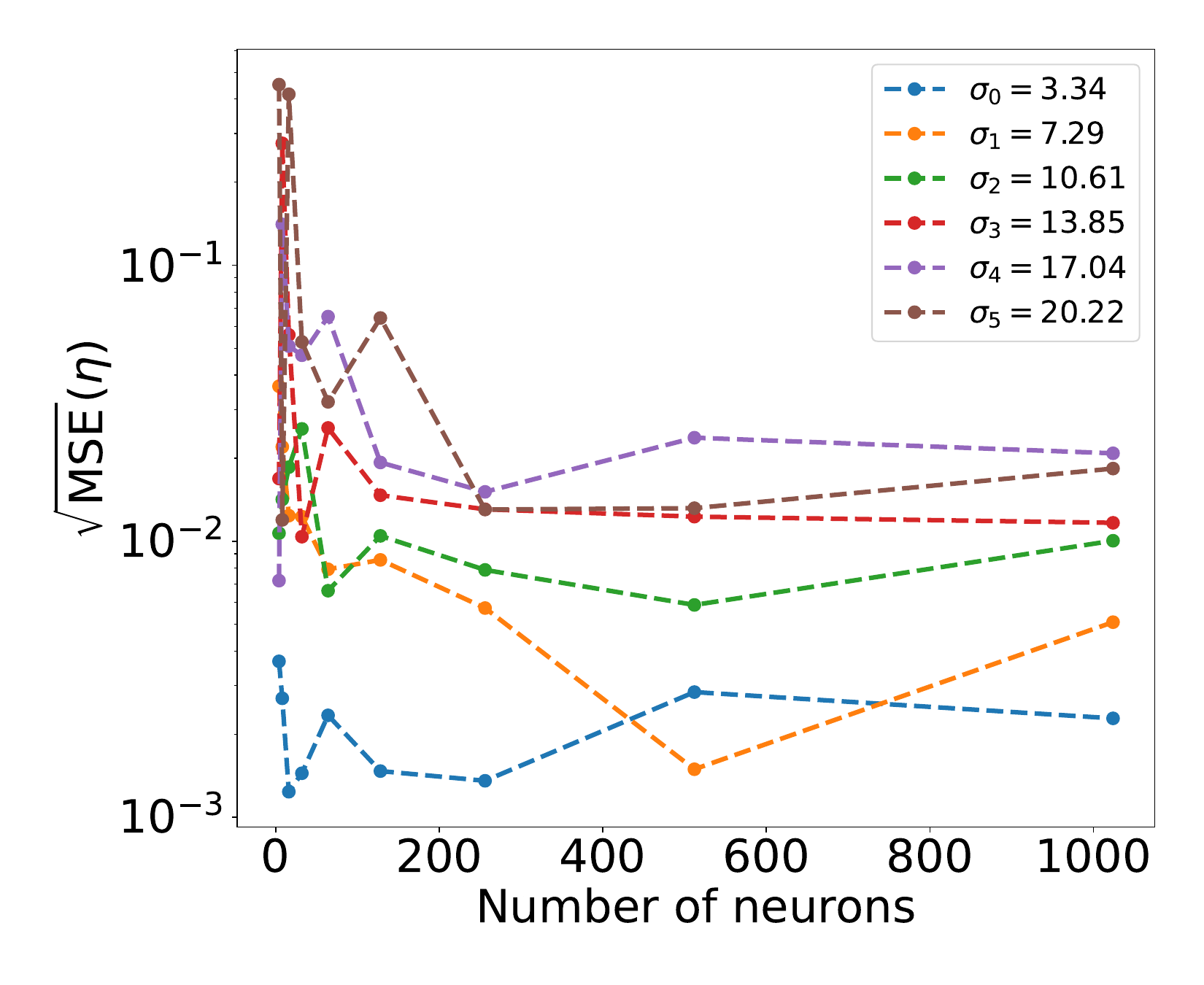}
   \includegraphics[width=0.49\linewidth]{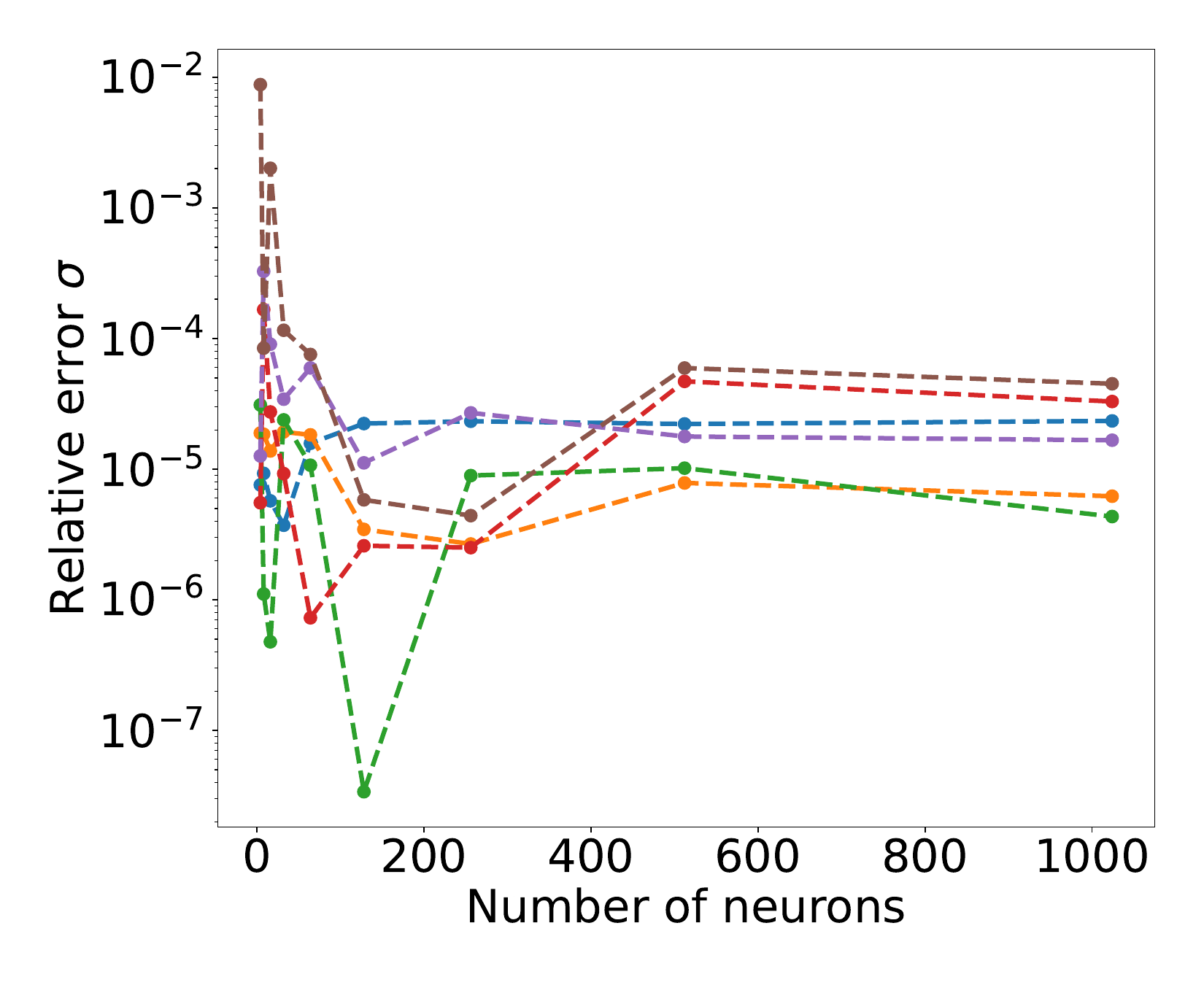}
   \caption{\justifying Upper figure: The square root of the mean square error of the eigenfunctions for all the eigenfrequencies for loss $10^{-7}$ is plotted with respect to the number of neurons for the f-PINN. Lower figure: The relative error of the frequencies for loss $10^{-7}$ for different number of neurons for the f-PINN.}
   \label{fig:errors_all_neurons}
\end{figure}

In addition, in Fig.~\ref{fig:errors_all_neurons} we show the impact of the number of neurons on the relative error of the frequency (lower panel) and the mean square error of the eigenfunctions, $\sqrt{{\rm MSE}(\eta_1) + {\rm MSE}(\eta_2)}$, (upper panel). The plots indicate that once the network has a sufficient number of neurons to be able to describe the function, then increasing their number does not improve the accuracy further. By analyzing these results, alongside the eigenfunction profiles, we can determine an appropriate number of neurons. As in the case of the plots with respect to the different values of the loss, Fig.~\ref{fig:errors_all_losses}, the MSE of $\eta_1$ is much larger than the one of $\eta_2$ and thus contributing the most to their sum. The relative error of the frequencies is around two order of magnitude lower than the MSE.

The selection of an activation function is crucial for PINNs. For the s-PINN we use the hyperbolic tangent, which is widely used in the literature. However, for the f-PINN, implementing this activation function led to inaccurate eigenfunctions, especially for higher-order modes. Thus, we explored other options, among them sigmoid, Leaky ReLU \cite{PyTorchActFunctions} and sinusoidal. The latter has the advantage that it is infinitely differentiable, and it was the only one that ensured the proper shape, as well as the exact number of nodes of the eigenfunctions.

Another important hyperparameter that needs to be tuned is the learning rate. We tested various learning rate values and monitored the number of epochs required for the PINN to reach the desired loss value. In addition, we checked whether the training was falling into a plateau and could not improve further or if it took many epochs to do so. We observed that for the s-PINN the best value for the learning rate was $10^{-4}$. On the contrary, for the f-PINN we implemented a learning rate scheduler, ReduceLROnPlateau. The scheduler reduces the learning rate by a factor of 0.5 when the loss does not show improvement in the previous $10^4$ epochs. The minimum learning rate is set to $5\cdot 10^{-7}$. The total number of epochs for both PINNs is of the order of $10^6$. This value is suitable as it is high enough so that the PINN is not forced to stop before finding a solution.

We use the Adam optimizer for both PINNs. We also tried the RMSprop and SGD \cite{PyTorchOptimizer}, but both of them required significantly more time to compute the first two eigenvalues, and were unable to recover higher-order frequencies even for a loss as low as $\varepsilon=10^{-8}$.

\section{Conclusions}
\label{sec:conclusions}
In this work, we revisit the eigenvalue problem associated to neutron star oscillations, in a simplified setup, and implement for the first time, in this context, a machine learning algorithm based on PINNs. We consider linear adiabatic perturbations of the Euler equations for a sphere of unit radius and constant density, pressure and sound speed. These equations can be expressed as a generalized eigenvalue problem, whose eigenvalues are the oscillation frequencies of the system.

PINNs have been used in the last years to solve eigenvalue problems linked to different fields of physics, including astrophysics.
The advantages of PINNs to other traditional methods, in terms of the absence of grids and the ease to impose physical constraints and boundary conditions, make them very promising to solve the eigenvalue problem associated to this problem. They are easy to implement through standard libraries.

 We construct an eigenvalue solver using PINNs in two acts. First, we perform a \textit{coarse search} over an (equally spaced) frequency domain using a PINN, which we name s-PINN. Its role is to identify the intervals that contain an eigenvalue by solving the system of equations, with boundary conditions at the origin, for distinct values of the frequency. For the second part, we introduce two independent methods to compute the eigenfrequency. In the first scheme, we use the same s-PINN to find the eigenvalue, applying a bisection algorithm. Alternatively, in the second method, we implement a specialized PINN, the f-PINN, for the calculation of the eigenvalue, where the eigenfrequency is a parameter of the network.

We present the first six eigenfrequencies and the corresponding eigenfunctions obtained by both methods; bisection and f-PINN. 
As the test case has analytic solutions, we can calculate the relative error of the frequencies and the mean square error of the eigenfunctions to the theoretical ones. Comparing the results, we show that although the f-PINN calculates the eigenfrequencies with much higher accuracy, it requires significantly more computational time. Specifically, the relative error of the frequencies computed by the bisection algorithm is of the order of $10^{-4}$, while those of the f-PINN are about two orders of magnitude lower. The MSE of the eigenfunctions is also slightly better in the f-PINN.

To test the performance of our code and its efficiency in identifying all the physical solutions, we thoroughly examined the convergence of each PINN. We show that the convergence depends mainly on the threshold value of the loss function. Additionally, we observe its impact on the eigenfunctions.
We establish the correlation between the loss value and the form of the radial profiles of the eigenfunctions. The MSE of the eigenfunctions becomes sufficiently low only for low values of the loss function. The importance of the value of the MSE is reflected on the shape of the eigenfunctions. If the MSE is high, then the eigenfunctions are recovered incorrectly.
That is evident, especially, for higher-order harmonics, where the complexity of the eigenfunctions cannot be captured, \eg the number of nodes is wrong. Moreover, we performed a thorough hyperparameter study for both networks in order to achieve the highest accuracy and the speed of the code.

One of the potential drawbacks of the method is that, even if the loss achieved is small, it cannot be guaranteed that the solution obtained is one of the eigenvalues. In principle, the algorithm could be converging to a local minimum of the loss, instead of the physical solution. For the example presented in this work, where we know the analytic solution, using either of the two methods proposed, we never encounter this problem. However, during the hyperparameter study, the PINN failed in some of the cases, and a possible cause could be the existence of local minima in the loss.

However, the main limitation is the computational time. When compared with the same test performed using the shooting algorithm \cite{Torres_et_al_I} or spectral methods \cite{2025Advection}, the computational time is about a factor $100$ longer for the case of PINNs, which make them non-competitive at this stage. Nevertheless, the efficiency of PINNs could be enhanced by incorporating innovative methods such as Fourier Features \cite{tancik:2020} or Adaptive Sampling \cite{Wu:2023, Zhao:2024}. Fourier Features transform input coordinates into a higher-dimensional space through randomized sinusoidal mappings, thereby enhancing the network’s capacity to capture high-frequency details. Adaptive Sampling improves both efficiency and accuracy by prioritizing training in regions with elevated residuals. Additionally, with the advent of quantum computers PINNs could become competitive (see e.g. \cite{2022arXiv220914754M,VADYALA2023100287}).

In this work, we demonstrate a successful application of PINNs as eigenvalue solvers in the simple scenario. We believe that this study can be extended to more complex cases with increased physical realism, which may have interest in neutron star asteroseismology. One of the applications is the study of perturbations of backgrounds without spherical symmetry, such as those present in rotating neutron stars (2D, axisymmetric) or in the presence of magnetic fields (3D, non-symmetric). Neural networks could be adapted easily to such multidimensional eigenvalue problems by simply increasing the number of inputs. What remains to be seen are the challenges that present the multidimensional, in terms of the requirements for the PINN architecture, the loss and the computational time.

%-------------------------------------------------------------- 

\section{Acknowledgments}
We would like to thank Prof. Arnau Rios for useful suggestions and productive discussions. This research was initiated during the participation at the Institute for Pure \& Applied Mathematics (IPAM), University of
California Los Angeles (UCLA) at the occasion of the Mathematical and Computational Challenges in the Era of Gravitational-Wave
Astronomy workshop. IPAM is partially
supported by the National Science Foundation through
award DMS-1925919. The authors would like to thank
IPAM and UCLA for their warm hospitality. This article is based upon work from COST Action CA17137, supported by COST (European Cooperation in Science and Technology).
The authors acknowledge the support from the grants 
Prometeo CIPROM/2022/49 from the Generalitat Valenciana, and  PID2021-125485NB-C21 from the Spanish Agencia Estatal de Investigación funded by MCIN/AEI/10.13039/501100011033 and ERDF A way of making Europe.

\bibliographystyle{iopart-num}
\bibliography{references}

\end{document}